\documentclass[groupedaddress,
  aip,
  jap,
  amsmath,amssymb,
  preprint,
]{revtex4-1}
\usepackage{graphicx,color}
\usepackage{amsmath}
\usepackage{natbib}
\usepackage{epsfig}
\begin{document}

\title{Practical dispersion relations for strongly coupled plasma fluids}

\author{Sergey A. Khrapak}
\email{Sergey.Khrapak@univ-amu.fr}
\affiliation{Aix-Marseille University, CNRS, PIIM, 13397 Marseille, France; Institut f\"ur Materialphysik im Weltraum, Deutsches Zentrum f\"ur Luft- und Raumfahrt (DLR), 82234 We{\ss}ling, Germany; Joint Institute for High Temperatures, Russian Academy of Sciences, 125412 Moscow, Russia}

\date{\today}

\begin{abstract}
Very simple explicit analytical expressions are discussed, which are able to describe the dispersion relations of longitudinal waves in strongly coupled plasma systems such as one-component plasma and weakly screened Yukawa fluids with a very good accuracy. Applications to other systems with soft pairwise interactions are briefly discussed.     
\end{abstract}

\maketitle

\section{Introduction}

Collective dynamics in strongly coupled plasmas is an important research topic with interdisciplinary relations (e.g. to collective motion in other condensed matter systems). It is particularly relevant for complex (dusty) plasmas -- systems of charged macroscopic particles immersed in a plasma environment. The charge of these particles is typically very high ($10^3-10^4$ elementary charges) and they usually form condensed liquid and solid phases due to strong electrical interactions. It is well understood that dispersion properties of strongly coupled plasmas significantly deviate from those of ideal gaseous plasmas.~\cite{FortovUFN,FortovPR,Bonitz2010} 

There is a number of different theoretical approaches to describe waves in strongly coupled systems that have been discussed in the context of complex plasmas. These include, for example, the approaches of generalized hydrodynamics,~\cite{Kaw1998,Kaw2001,Diaw2015} multicomponent kinetic approach,~\cite{Murillo2000} and the quasilocalized charge approximation (QLCA).~\cite{GoldenPoP2000,RosenbergPRE1997,KalmanPRL2000} Comparison with direct numerical simulations documented good performance of the QLCA model, at least for weakly and moderately screened systems (one-component plasma and Yukawa fluids with interparticle separation equal to several screening lengths or shorter).~\cite{KalmanPRL2000,DonkoJPCM2008,OhtaPRL2000,HamaguchiPS2001} In this paper we demonstrate that the QLCA model can be reduced to a set of two simple coupled explicit expressions, which allow to describe very accurately the longitudinal dispersion relations in a wide parameter regime.    

The QLCA model [also known as quasi-crystalline approximation (QCA)~\cite{Hubbard1969,Fingerprints}] relates wave dispersion relations to the interparticle interaction potential and the equilibrium radial distribution function (RDF) $g(r)$, characterizing structural properties of the system. It can be considered as either a generalization of the random phase approximation or as a generalization of the phonon theory of solids~\cite{Hubbard1969} (this is why the term QCA is appropriate~\cite{Takeno1971}). The radial distribution function in the fluid phase can in principle be obtained from direct numerical simulations or from integral equations of liquid state theory. For classical crystals with isotropic interactions, in addition to numerical sumulations, a shortest-graph method has been recently applied,~\cite{YurchenkoJCP2014,YurchenkoJCP2015} which can be further modified to include anharmonicity effects.~\cite{YurchenkoJPCM2016}  It turns out, however, that to describe the long-wavelength portions of the dispersion curves a very accurate knowledge of $g(r)$ is unnecessary. The main idea is that since in the QLCA model the function $g(r)$ appears under the integral, an appropriate model for $g(r)$, even if it does not describe very accurately the actual structural properties of the system, can nevertheless be helpful in estimating the behaviour of dispersion curves. 

For sufficiently steep interactions (like for instance the Lennard-Jones potential) the radial distribution function can be effectively modelled by a delta function $g(r)\simeq {\mathcal A}\delta(r-r_0)$, where $r_0$ is roughly the mean interparticle separation and ${\mathcal A}$ is a properly adjusted weight.~\cite{Hubbard1969} It was demonstrated previously that this is a very useful approximation to describe wave dispersion of conventional atomic liquids (e.g. such as Ar and Rb).~\cite{Copley1975}   

The situation is, however, different for soft interactions occurring in the plasma-related context. Here, due to a long-range character of the interaction potential, the function appearing under the integral of the QLCA model is also long-ranged and a different motel for $g(r)$ seems more appropriate. Recently, it has been proposed~\cite{KhrapakPoP2016} to take the simplest possible model $g(r)$ of the form   
\begin{equation}\label{gOFr}
 g(r)=\theta(r-R),
\end{equation}
where $\theta(x)$ is the Heaviside step function and $R$ is again of the order of the mean interparticle separation. Physically, this trial form seems reasonable, because the main contribution to the long-wavelength dispersion corresponds to long length scales, where $g(r)\simeq 1$. The excluded volume effect for $r\leq R$ allows to properly account for strong coupling. Previously, a similar RDF was employed to analyse the main tendencies in the behaviour of specific heat of liquids and dense gases at low temperatures.~\cite{Stishov} It was also used to calculate the dispersion relation of Coulomb bilayers and superlattices at strong coupling.~\cite{Golden1993} The approach was demonstrated to be satisfactory even without precise determination of $R$. However, the radius of the correlational hole $R$ is generally not a free parameter of the approximation. It was proposed~\cite{KhrapakPoP2016} to determine $R$ from the condition that the model form (\ref{gOFr}) delivers good accuracy for the excess internal energy and pressure (which can also be expressed as integrals over $g(r)$ for pairwise interactions~\cite{HansenBook,MarchBook}). This simple approximation demonstrated very good accuracy when applied to weakly screened Yukawa systems.~\cite{KhrapakPoP2016} Here we go further and provide an explicit expression for the excluded hole radius $R$. We then demonstrate that an emerging set of two simple coupled expressions allows us to describe very accurately the long-wavelength dispersion relations of strongly coupled plasma fluids. To do this, detailed comparison between the calculated dispersion relations and the benchmark results from previous numerical simulations is performed. 

\section{Methods}

The Yukawa systems considered here are characterized by the repulsive interaction potential of the form $V(r)=(Q^2/r)\exp(-r/\lambda)$, where $Q$ is the particle charge, $\lambda$ is the screening length, and $r$ is the distance between a pair of particles. The phase state of the system is conventionally described  by the two dimensionless parameters,~\cite{HamaguchiPRE1997} which are the coupling parameter $\Gamma=Q^2/aT$, and the screening parameter $\kappa=a/\lambda$. Here $T$ is the system temperature (in energy units), $n$ is the particle density, and $a=(4\pi n/3)^{-1/3}$ is the Wigner-Seitz radius (three-dimensional systems are considered here). When $\kappa\rightarrow 0$ the one-component plasma (OCP) limit is recovered, however neutralizing background should be added to keep thermodynamic quantities finite. Yukawa systems are normally referred to as strongly coupled when $\Gamma\gg 1$. The Yukawa potential is considered as a reasonable starting point to model interactions in complex (dusty) plasmas and colloidal dispersions,~\cite{FortovPR,IvlevBook} although in many cases the actual interactions (in particular, their long-range asymptotes) are much more complex.~\cite{FortovPR,KhrapakCPP2009} The effects of the long-range asymptote of the interaction (deviations from the pure Yukawa shape) on the dispersion relations of the longitudinal waves in complex plasmas have been recently discussed.~\cite{Fingerprints}

\begin{figure}
\includegraphics[width=10cm]{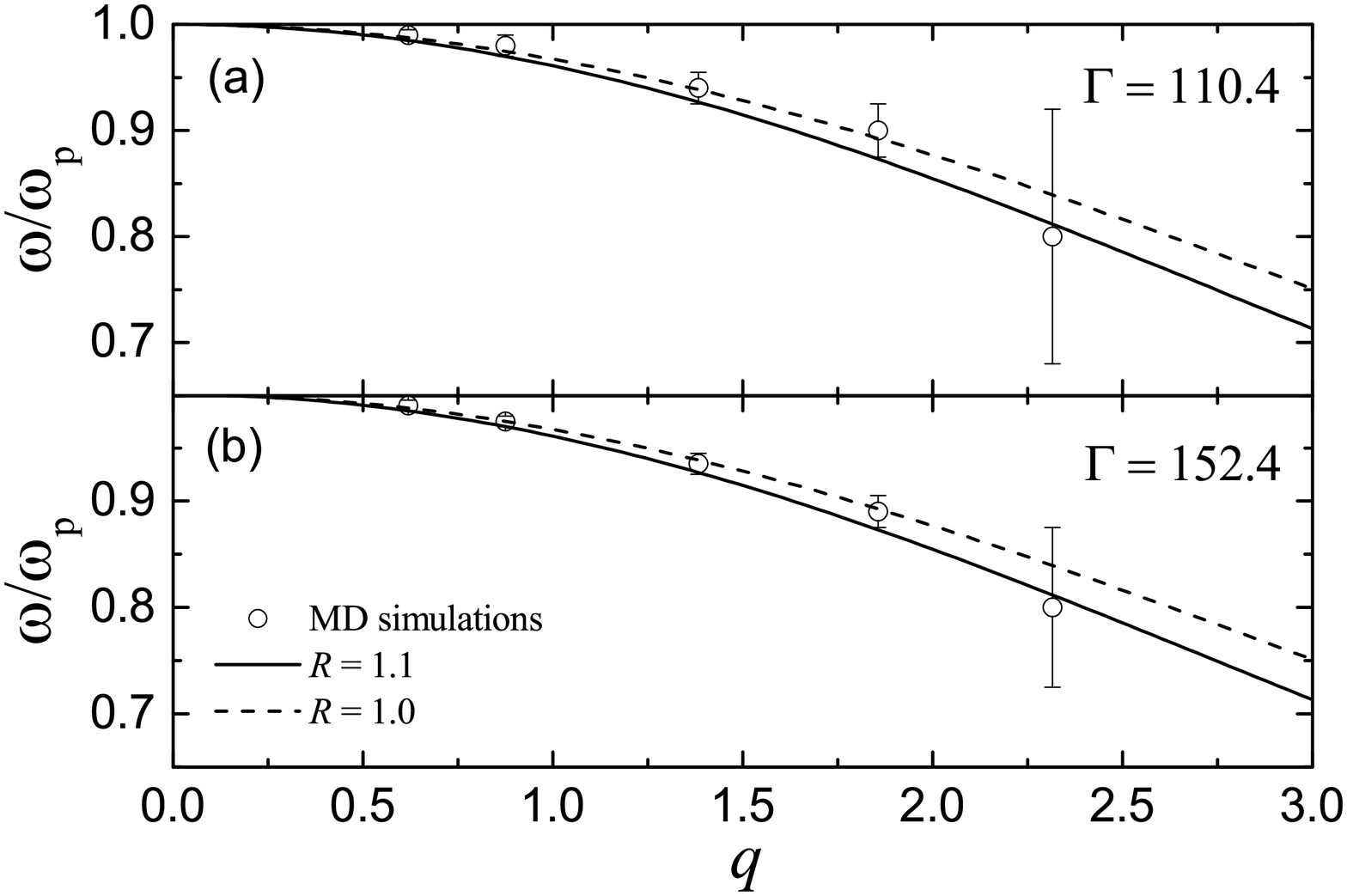}
\caption{Dispersion of the longitudinal (plasmon) mode of the strongly coupled  OCP. The coupling parameter is $\Gamma= 110.4$ (a) and $\Gamma= 152.4$ (b). Symbols correspond to the results from MD simulations.~\cite{HansenPRL1974} The curves are plotted using Eq.~(\ref{OCPdisp}) with two different values of the parameter $R$ ($R=1.1$-solid; $R=1.0$-dashed).   }
\label{Fig1}
\end{figure}

For the Yukawa interaction potential and RDF model of Eq. (\ref{gOFr}), the generic expressions for the QLCA dispersion relations (containing rather complex integrals~\cite{KalmanPRL2000,DonkoJPCM2008}) can be integrated analytically. The resulting dispersion relation of the longitudinal mode is~\cite{KhrapakPoP2016}
\begin{equation}\label{L1}
\omega^2=\omega_{\rm p}^2e^{-R\kappa}\left[\left(1+R\kappa\right)\left(\frac{1}{3}-\frac{2\cos Rq}{R^2q^2}+\frac{2\sin Rq}{R^3q^3} \right) - \frac{\kappa^2}{\kappa^2+q^2}\left(\cos Rq+\frac{\kappa}{q}\sin Rq \right)\right],
\end{equation}  
where $\omega$ is the frequency, $q=ka$ is the reduced wave number, $\omega_{\rm p}=\sqrt{4\pi Q^2 n/m}$ is the plasma frequency scale, $m$ is the particle mass, and $R$ is expressed in units of $a$. In the limit where correlations are absent, $R\rightarrow 0$, the conventional dust-acoustic wave dispersion relation~\cite{RaoDAW} is recovered
\begin{equation}
\omega^2=\frac{\omega_{\rm p}^2q^2}{\kappa^2+q^2}.
\end{equation}
In the strongly coupled fluid regime we need to evaluate the appropriate radius of the correlational hole, $R$. As pointed out above, a reasonable approach is to require that the system excess energy and pressure are reproduced accurately when the approximation (\ref{gOFr}) is substituted in the corresponding integral equations. The difference between the energy and pressure routes is very subtle for weakly screened Yukawa systems,~\cite{KhrapakPoP2016} and we use the energy route here, which proves to be somewhat more simple. For soft interactions considered here, it is well known that the thermodynamic quantities such as internal energy, pressure, or compressibility are dominated by the static contribution in the regime of strong coupling.~\cite{KhrapakJCP2015} For Yukawa systems this  static contribution is very well accounted for by the ion sphere model (ISM), provided $\kappa$ is not very large.~\cite{RosenfeldPRE2000,KhrapakISM} The excess internal energy of the ISM model is
\begin{equation}\label{ISM}
u_{\rm ex}=\frac{\kappa(\kappa+1)\Gamma}{(\kappa+1)+(\kappa-1)e^{2\kappa}}.
\end{equation}
On the other hand, the integral energy equation, for the model RDF of Eq.~(\ref{gOFr}), yields
\begin{equation}\label{energy}
u_{\rm ex}=\frac{3\Gamma}{2\kappa^2}\left(1+R\kappa\right)e^{-R\kappa}.
\end{equation}
In equating (\ref{ISM}) and (\ref{energy}) we use the fact that $R\simeq 1$, so that we can substitute $(1+R\kappa)$ with $(1+\kappa)$ in Eq.~(\ref{energy}). In doing so we get a particularly simple explicit expression for the excluded volume radius $R$, 
\begin{equation}\label{R}
R(\kappa)\simeq 1+\frac{1}{\kappa}\ln \left[\frac{3 \cosh (\kappa)}{\kappa^2}-\frac{3 \sinh
(\kappa)}{\kappa^3}\right].
\end{equation}
Roughly, this yields $R(\kappa)\simeq 1+\kappa/10$ for the regime of weak screening considered. Note that the size of the correlational hole turns out to be virtually independent of $\Gamma$. This indicates that in the regime of strong coupling wave dispersion relations (in properly normalized units) are practically insensitive to the exact coupling strength. The set of equations (\ref{L1}) and (\ref{R}) represents a very simple practical tool to describe the dispersion relations of the longitudinal waves in strongly coupled Yukawa systems. Below, we demonstrate the high accuracy of this approximation at long wavelengths by comparing it with the results from benchmark numerical simulations. 

\section{Results and discussion}    

\begin{figure*}
\includegraphics[width=15cm]{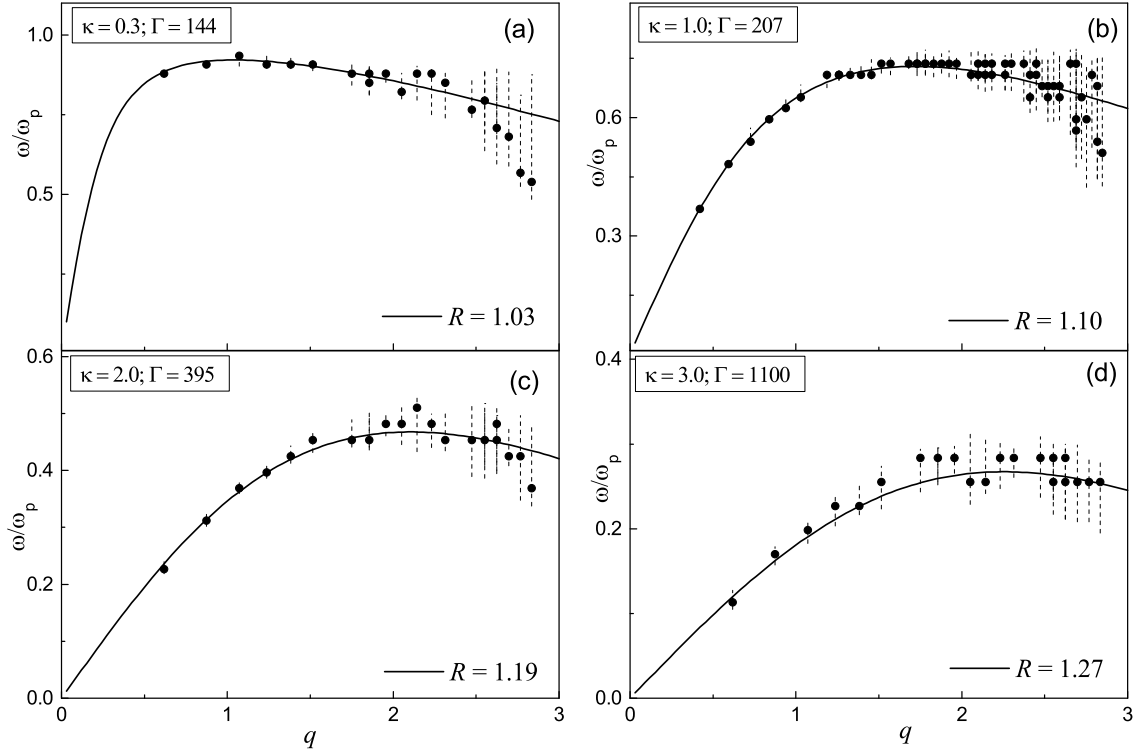}
\caption{Dispersion of the longitudinal waves in Yukawa fluids near the fluid-solid phase transition (the values of $\kappa$ and $\Gamma$ are indicated in the top left corners of Figs.~(a)-(d)). Symbols with error bars correspond to the results from numerical simulations.~\cite{OhtaPRL2000,HamaguchiPS2001} Solid curves are calculated using Eqs.~(\ref{L1}) and (\ref{R}) from this paper. The corresponding values of $R$ appear in the bottom right corners (note, that to a good accuracy $R\simeq 1+\kappa/10$ ).}
\label{Fig2}
\end{figure*}

We start with the OCP limit, corresponding to the unscreened Coulomb interaction between the particles. The dispersion relation of the longitudinal plasmon mode follows directly from Eq.~(\ref{L1}) by taking the limit $\kappa\rightarrow 0$:
\begin{equation}\label{OCPdisp}
\omega^2=\omega_{\rm p}^2\left(\frac{1}{3}-\frac{2\cos Rq}{R^2q^2}+\frac{2\sin Rq}{R^3q^3} \right).
\end{equation}
An approximate equation (\ref{R}) yields $R=1$ in this limit. A somewhat more accurate analysis, which takes into account specifics of the OCP, results in a very close value of $R= \sqrt{6/5}\simeq 1.09545$.~\cite{KhrapakPoP2016} The plasmon dispersion relations calculated with the help of Eq.~(\ref{OCPdisp}) are plotted in Fig.~\ref{Fig1} for the two strongly coupled state points. The symbols correspond to the results derived from MD simulations.~\cite{HansenPRL1974} The agreement is very good. 

The accuracy documented in Fig.~\ref{Fig1} is not expected to hold at weaker coupling. This is because the original QLCA model is itself not designed for this regime~\cite{GoldenPoP2000} and hence lacks accuracy.  In particular, it cannot describe the transition from the positive to negative dispersion~\cite{MithenAIP2012,HansenJPL1981} at $\Gamma\simeq 10$. Here positive/negative dispersion refers to the positive/negative sign of $d\omega/d q$ at $q\rightarrow 0$ (note, that in this sense the dispersion is always negative within the QLCA model). Useful modifications which allow to capture the onset of negative dispersion at moderate coupling have been recently discussed.~\cite{KhrapakNegative}  

Next, we compare the results of calculation using Eqs.~(\ref{L1}) and (\ref{R}) with the results obtained for the dispersion of Yukawa fluids using molecular dynamics simulations.~\cite{OhtaPRL2000,HamaguchiPS2001}
This comparison is shown in Fig.~\ref{Fig2} for four state points, located near the fluid-solid (freezing) curve. In all cases the agreement is impressive in the range $q\lesssim 3$, especially taking into account the level of simplifications involved.     

\begin{figure}
\includegraphics[width=10cm]{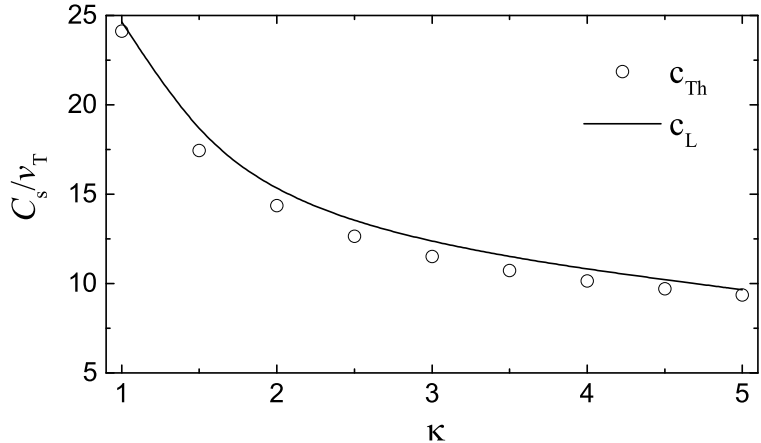}
\caption{Sound velocity of Yukawa systems at the melting temperature (Yukawa melts). The sound velocity, normalized by the thermal velocity $v_{\rm T}=\sqrt{T/m}$, is plotted as a function of the screening parameter $\kappa$. The symbols correspond to the thermodynamic definition of the adiabatic sound velocity.~\cite{KhrapakPRE2015_Sound} The solid curve corresponds to the elastic sound velocity calculated from Eq.~(\ref{Sound}). Note that the sound velocity diverges in the OCP limit ($\kappa\rightarrow 0$).  }
\label{Fig3}
\end{figure}

In the long-wavelength limit ($q\rightarrow 0$), Eq.~(\ref{L1}) provides the QLCA elastic longitudinal sound velocity:
\begin{equation}\label{Sound}
c_{\rm L}^2=\frac{\exp(-\kappa R)}{\kappa^2}\left(1+\kappa R+\tfrac{13}{30}\kappa^2 R^2 + \tfrac{1}{10}\kappa^3 R^3\right),
\end{equation}
where the velocity is expressed in units of $\omega_{\rm p}a$ (to get the sound velocity in units of thermal velocity, $v_{\rm T}=\sqrt{T/m}$, one should multiply $c_{\rm L}$ by the factor $\sqrt{3\Gamma}$). We plot the resulting QLCA sound velocity for Yukawa systems at the fluid-solid phase transition (Yukawa melts) in Figure~\ref{Fig3}. The curve is obtained using Eq.~(\ref{Sound}) with the coupling parameters at melting tabulated previously.~\cite{HamaguchiPRE1997} The symbols correspond to the conventional thermodynamic approach~\cite{KhrapakPRE2015_Sound,KhrapakPPCF2016} to the adiabatic sound velocity in fluids. It is observed, that the QLCA elastic sound velocity $c_{\rm L}$ slightly overestimates the thermodynamic sound velocity $c_{\rm Th}$. This observation is not surprising~\cite{KhrapakPRE2015_Sound,KhrapakPoP2016_Relations} and the reason for this overestimation is well understood. For systems with soft pairwise interactions in the strongly coupling regime, the thermodynamic sound velocity is to a good accuracy related to the longitudinal ($c_{\rm L}$) and transverse ($c_{\rm T}$) elastic velocities via $c_{\rm Th}^2\simeq c_{\rm L}^2-\tfrac{4}{3}c_{\rm T}^2$.~\cite{KhrapakPoP2016_Relations,LL,Trachenko2015} Since normally $c_{\rm L}\gg c_{\rm T}$ in this regime, the QLCA longitudinal elastic sound velocity $c_{\rm L}$ is close, but slightly above the thermodynamic sound velocity $c_{\rm Th}$. As the potential steepness increases (the regime not considered here), the difference between $c_{\rm L}$ and $c_{\rm Th}$ will also increase. It has been shown recently that in the limit of hard-sphere-like interactions, QLCA becomes grossly inaccurate and should not be applied.~\cite{KhrapakJCP2016,KhrapakSciRep2017}  In particular, this happens when the ratio of potential-to-kinetic energy drops below unity even for dense fluids in the vicinity of the fluid-solid phase transition.~\cite{BrazhkinPRE2012} In this regime fluids exhibit strong correlations in the absence of strong coupling (extreme example corresponds to hard-sphere fluids where potential interactions are absent at all).~\cite{KhrapakSciRep2017}   The conventional plasma fluids considered here are normal in this sense: Strong correlations are    
always related to strong coupling. 

In the short-wavelength limit ($q\rightarrow \infty$), Eq.~(\ref{L1}) tends to the QLCA Einstein frequency,
\begin{equation}\label{Einstein}
\Omega_{\rm E}^2=\frac{\omega_{\rm p}^2}{3}e^{-R\kappa}\left(1+R\kappa\right),
\end{equation}
which is the oscillation frequency of a single particle in the fixed environment of other particles, characterized by a given RDF. 
Combining Eqs. (\ref{Einstein}) ans (\ref{energy}) we immediately arrive at
\begin{displaymath}
\left(\frac{\Omega_{\rm E}}{\omega_{\rm p}}\right)^2=\frac{2\kappa^2 u_{\rm ex}}{9\Gamma}.
\end{displaymath}
This represents the {\it exact} relation between the Einstein
frequency and reduced excess energy in the special case of the Yukawa interaction potential. Thus, the approach is virtually exact in the
short-wavelength limit.

At intermediate wavelengths (i.e. at $3\lesssim q\lesssim \infty$) the deviations between QLCA calculations with exact and model RDF are significant.~\cite{KhrapakPoP2016} This regime is, however, of rather limited interest from the practical point of view.   
  
\section{Conclusion}
  
The main results from this study are as follows. Simple approximation for the radial distribution function has been used to derive analytical expressions for the wave dispersion in strongly coupled plasma fluids 
within the framework of the QLCA model. The performance of this approximation has been tested against the benchmark results from previous numerical simulations and a very good agreement has been documented at long wavelengths (the approximation is also virtually exact in the short-wavelength limit). This agreement is unlikely related to the particular form of the interaction  (Yukawa and Coulomb) considered here. Rather, the approach is expected to provide accurate and reliable results also for other related soft pairwise interactions. A recent supporting example is given by the two-dimensional one-component-plasma with logarithmic interactions, where the approach works extremely well.~\cite{2DOCP} Thus, a new simple and accurate tool to describe collective dynamics in soft interacting particle systems, without accurate knowledge of structural details, emerges.

\begin{acknowledgments} 
This work was supported by the A*MIDEX project (Nr.~ANR-11-IDEX-0001-02) funded by the French Government ``Investissements d'Avenir'' program managed by the French National Research Agency (ANR).
\end{acknowledgments}

\bibliographystyle{aipnum4-1}
\bibliography{SC_References}

\end{document}